# Measurement of Positron Showers with a Digital Hadron Calorimeter

Burak Bilki[d], John Butler[b], Ed May[a], Georgios Mavromanolakis[c,1], Edwin Norbeck[d], José Repond[a], David Underwood[a], Lei Xia[a], Qingmin Zhang[a,2]

[a]*Argonne National Laboratory, 9700 S. Cass Avenue, Argonne, IL 60439, U.S.A.*
[b]*Boston University, 590 Commonwealth Avenue, Boston, MA 02215, U.S.A.*
[c]*Fermilab, P.O. Box 500, Batavia, IL 60510-0500, U.S.A.*
[d]*University of Iowa, Iowa City, IA 52242-1479, U.S.A.*

**Abstract.** A small prototype of a finely granulated digital hadron calorimeter with Resistive Plate Chambers as active elements was exposed to positrons of 1 – 16 GeV energy from the Fermilab test beam. The response function, energy resolution, as well as measurements of the shape of electromagnetic showers are presented. The data are compared to a Monte Carlo simulation of the set-up.

**Keywords:** Calorimetry, Linear Collider, Particle Flow Algorithms, Resistive Plate Chambers.

**PACS:** 29.40.Vj, 29.40.Cs, 29.40.Gx

## INTRODUCTION

Particle Flow Algorithms (PFAs) attempt to measure all particles (originating from the interaction point of a typical colliding beam detector) in a jet individually, using the detector component providing the best momentum/energy resolution [1,2]. In this context and in preparation for the construction of a lager calorimeter module, a small prototype of a finely granulated hadron calorimeter (HCAL) using Resistive Plate Chambers (RPCs) as active elements was assembled. The prototype featured 1 x 1 cm$^2$ readout pads and a total of 1536 channels in six layers, interleaved with absorber plates. The readout system applied a single threshold to each pad (corresponding to a 1-bit resolution), hence the designation of Digital Hadron Calorimeter (DHCAL). The stack was exposed to positrons of the Fermilab test beam in the 1 – 16 GeV energy range. Measurements of the response function, the energy resolution and the shower shapes are presented and compared to expectations from Monte Carlo simulations based on GEANT4 [3] and a standalone program modeling the response of RPCs.

---

[1] Also affiliated with University of Cambridge, Cavendish Laboratory, Cambridge CB3 OHE, U.K.
[2] Also affiliated with Institute of High Energy Physics, Chinese Academy of Sciences, Beijing 100049, China and Graduate University of the Chinese Academy of Sciences, Beijing 100049,China.

This research was performed within the framework of the CALICE collaboration [4], which develops imaging calorimetry for the application of PFAs to the measurement of hadronic jets at a future lepton collider.

## DESCRIPTION OF THE CALORIMETER STACK

The calorimeter stack consisted of six chambers interleaved with absorber plates containing 16 mm thick steel and 4 mm thick copper and corresponding to approximately 1.2 radiation lengths each. The chambers measured 20 x 20 cm$^2$ and featured two glass plates. The thickness of the glass plates was 1.1 mm and the gas gap was maintained with fishing lines with a diameter of 1.2 mm.

The chambers were operated in avalanche mode with a high voltage setting of 6.3 kV. The gas consisted of a mixture of three components: R134A (94.5%), isobutane (5.0%) and sulfur-hexafluoride (0.5%). For more details on the design and performance of the chambers, see [5,6].

The chambers were mounted on the absorber plates and these in turn were inserted into a hanging file structure. The gap between absorber plates was 13.4 mm, where 8.3 mm were taken by the chambers and their readout boards.

The electronic readout system was optimized for the readout of large numbers of channels. In order to avoid an unnecessary complexity of the system, the charge resolution of individual pads was reduced to a single bit (digital readout). The readout system consisted of several parts: the pad-boards covering an area of 16 x 16 cm$^2$, the front-end board, the front-end Application Specific Integrated Circuits (the so-called DCAL chips), the data concentrator and collector modules, and the timing and triggering module. For more details on the readout system see ref. [7].

Every layer contained 256 individual readout pads, each with an area of 1 x 1 cm$^2$. The entire stack counted 1536 readout channels, of which only ten appeared to be dead and provided no signal. A photograph of the calorimeter stack in the test beam is shown in Fig. 1.

## TEST BEAM SETUP AND DATA COLLECTION

The stack was exposed to positrons from the test beam at the Meson Test Beam Facility (MTBF) of Fermilab [8]. Positrons were produced with an upstream target and were momentum selected in the range between 1 and 16 GeV/c. The beam came in spills of 3.5 second length every one minute.

The readout of the stack was triggered by the triple coincidence of two large scintillator paddles, each with an area of 19 x 19 cm$^2$, located approximately 2.0 and 0.5 meters upstream of the stack, and an upstream Čerenkov counter. The latter selected positrons and efficiently rejected other particles in the beam, such as muons, pions or protons. Table I lists the number of triggers collected at each momentum setting together with the average beam intensity during a spill and the estimated fraction of positrons in the beam.

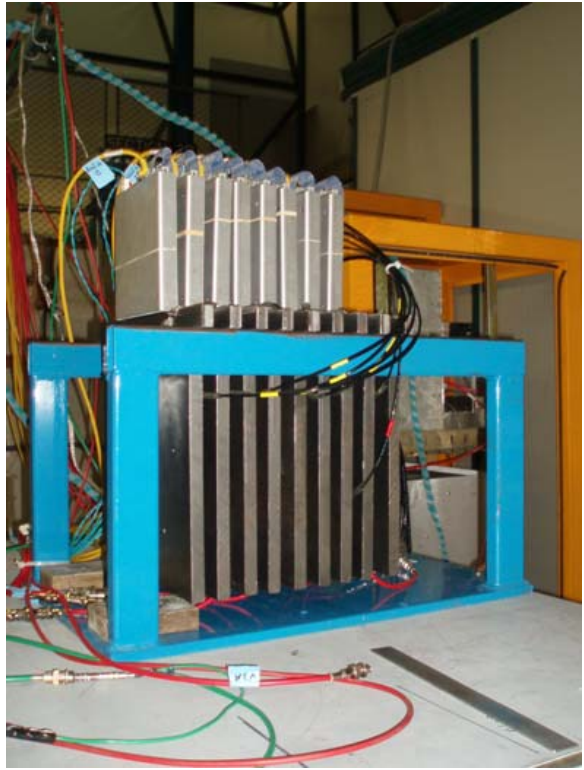

**Figure 1**. Photograph of the hanging file structure containing eight layers. The present measurements only utilized the first six of these layers.

| Momentum [GeV/c] | Number of events | Beam intensity [Hz] | Fraction of positrons [%] | Average number of points | Rate of points [Hz] |
|---|---|---|---|---|---|
| 1 | 10599 | 586 | 87 | 5.3 | 2695 |
| 2 | 8544 | 240 | 86 | 9.8 | 2030 |
| 4 | 13061 | 243 | 78 | 17.7 | 3338 |
| 8 | 39376 | 314 | 21 | 33.2 | 2182 |
| 16 | 6540 | 304 | 10 | 58.5 | 1822 |

**Table I.** Summary of the positron runs. The number of *points* refers to the number of energy deposits generated in the Monte Carlo simulation of the calorimeter stack. The average number of *points* is quoted for the layer containing the shower maximum and is per generated positron. The last column refers to the estimated rate of *points* at shower maximum from positrons taking into account the beam intensity and the fraction of positrons in the beam.

At the face of the stack the beam spot for momenta between 4 and 16 GeV/c was somewhat collimated with a sigma of approximately 2 cm both horizontally and vertically. At the lowest two energies the beam spot appeared to cover the entire the readout area of the chambers.

Figure 2 shows various views/projections of a typical event with an 8 GeV/c positron induced shower. Notice the high density of hits.

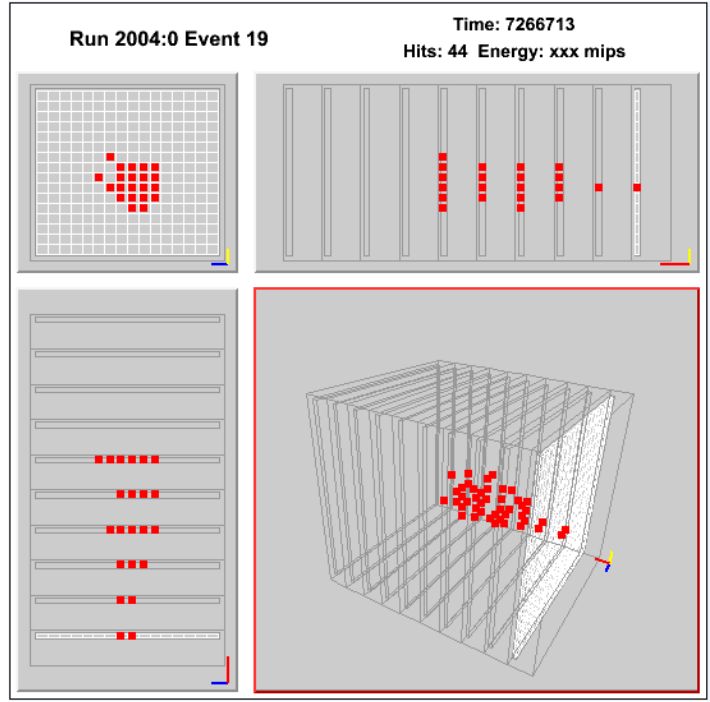

**Figure 2.** Event display of a positron induced shower.

## MONTE CARLO SIMULATION AND CALIBRATION

The test beam set-up has been simulated with a Monte Carlo program based on the GEANT4 package [3] and a standalone program RPCSIM for the simulation of the response of RPCs. The GEANT4 simulation includes the relevant portions of the beam line, the trigger counters[3], and the details of the calorimeter stack. The range cut in the GEANT4 program was left at its default value of 1.0 mm. Positron data are generated at the various energies utilized in the test beam and with their corresponding lateral beam profiles. The spatial coordinates of any energy deposition in the gas gap of an RPC is recorded for further analysis. In the following these energy depositions are named *points*.

---

[3] The 16 GeV data showed an unusually large number of early showers. This data was taken before the data at other energies and before a re-arrangement and simplification of the beam line. Additional material (1/4 $X_0$ of iron) was introduced into the simulation of the beam line in order to obtain a satisfactory simulation of the measured longitudinal shower profile.

For each generated *point*, the RPCSIM program generates a signal charge Q, distributes this charge over the pads, sums up all charges on a given pad, and applies a threshold *T* to identify the pads with hits.

The signal charges are generated according to the measured [5] spectrum of avalanche charges utilizing cosmic rays. The latter is fitted to the following functional form:

$$N(Q) = \alpha Q^\beta e^{-\gamma Q}, \tag{1}$$

where the parameters α, β and γ depend on the operating high voltage. An additional parameter $Q_0$ was introduced to accommodate possible differences between the charge distributions as measured in the laboratory and obtained in the test beam set-up:

$$Q' = Q + Q_0 \tag{2}$$

As a function of lateral distance R to a given *point* the induced charge in the plane of the readout pads is assumed to decrease exponentially [9] with a slope *a*:

$$dQ/dR = (Q'/a) \cdot e^{-Ra}$$

The three parameters $Q_0$, *T*, and *a* were tuned to reproduce the distribution of hits in individual chambers and the sum of hits in six consecutive chambers as measured in a broadband muon beam [6]. Figure 3 shows both distributions and the result of the simulation after tuning of the parameters. Whereas the hit distribution in single chambers could be simulated adequately, the distribution of the sums shows some differences between data and simulation, possibly due to minor differences in the angular distribution of the muons in the test beam and in the simulation. It was found that these differences could not be reduced without simultaneously degrading the simulation of the response in individual layers. The best values of the simulation parameters were found to be $Q_0$ = -0.2 pC, *T* = 0.60 pC and *a* = 0.17 cm.

An additional parameter was introduced to simulate a possible inefficiency of the chambers for additional *points* within a radius $d_{cut}$ from a given *point*. Due to the low number of *points* in muon induced events this parameter could not be tuned using the muon data, but had to be tuned with the positron data itself. It was found that a value of $d_{cut}$ = 0.1 cm best reproduced the average sum of the hit distributions in the calorimeter.

No attempt was made to simulate possible inefficiencies of the chambers [10] due to the high particle flux in positron induced showers. The performance of the individual chambers was similar enough rendering any chamber-specific fine-tuning of the calibration redundant.

The Monte Carlo generated events were formatted the same way as the test beam data and were analyzed by identical programs.

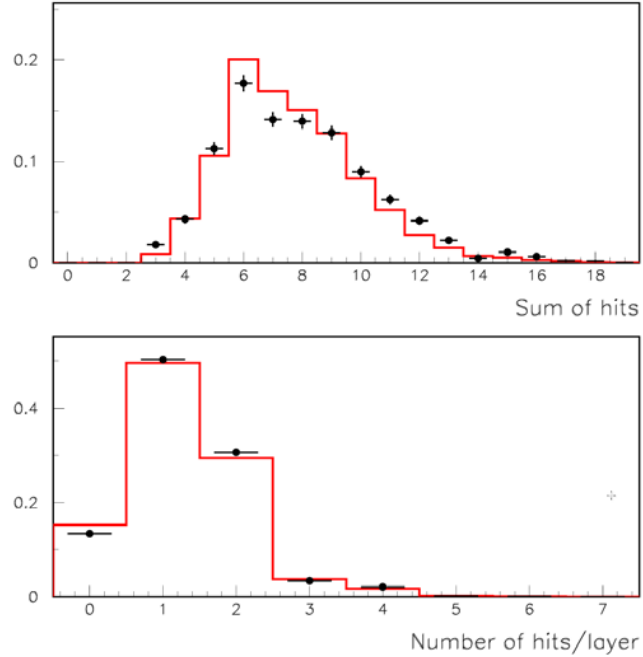

**Figure 3.** Normalized distribution of the sum of hits in six layers (top) and the distribution of hits averaged over six layers (bottom) in a broadband muon beam. The red histograms represent the results of a GEANT4 simulation of the set-up together with a standalone simulation of the response of RPCs.

## EVENT SELECTION

The event selection insured the high purity of the positron data, while rejecting muons, hadrons and multi-particle events. The selection criteria are described in the following:

a) *Requirement of at least one layer with a minimum of three hits.* This cut reduces the contamination from accidental triggers and possible backgrounds from muon induced events.
b) *Requirement of at least four layers with hits.* This cut removed partly contained events and reduced contamination from accidental triggers.
c) *Requirement of exactly one cluster of hits in the first layer.* This cut removed events with more than one particle entering the calorimeter or with showers which initiated upstream of the calorimeter stack. Clusters of hits were reconstructed as aggregates of cells with at least one side in common between two cells.
d) *Requirement of no more than six hits in the first layer.* This cut removed events with electromagnetic showers which initiated upstream of the calorimeter.
e) *Fiducial cut on the position of the cluster in the first layer.* In order to contain the showers laterally, the cluster in the first layer was required to be at least 3 cm from the edge of the readout area of the chambers. However, due to the high particle flux in electromagnetic showers, the data at 4, 8, and 16 GeV was plagued with inefficiencies of the chambers placed close to the shower maxima. For these data, the fiducial cut excluded an area of 3 x 3 cm$^2$ at the center of the chambers, while extending the fiducial region up to 1 cm from the edge of the readout area. The observed loss in efficiency was compatible with expectations based on the measured rate capability of the chambers [10], the intensity, lateral profile and particle composition of the beam and the predicted number of avalanches in a given layer for positron (see Table I.) and pion induced events. Due to the locality of the rate induced inefficiencies over the surface of a chamber [10] the largest loss of efficiency was expected at 8 GeV, where the beam spot was found to be relatively narrow.

f) *Cut on the spill time for the 8 GeV data.* To reduce the effect of high particle fluxes on the efficiency of the chambers, only events in the first 0.2 seconds of a given spill were retained. This cut was only applied to the 8 GeV data where sufficient statistics was available.

## RESPONSE AND RESOLUTION

With a digital hadron calorimeter the energy of an incoming hadron, $E_{hadron}$, can be reconstructed from the number of hits associated with that particle. Ignoring effects of high-density sub-clusters, which might require non-linear corrections, the sum of hits is expected to be proportional to $E_{hadron}$. This paper investigates the response of the calorimeter to positrons of varying energy. Due to the high density of the associated electromagnetic showers, the response is not expected to be linear with energy.

To illustrate the development of electromagnetic showers in this type of calorimeter, Figs. 4 and 5 show the distribution of the number of hits in each layer individually for 2 and 8 GeV positrons, respectively. The distributions are compared to the prediction of the Monte Carlo simulation and are seen to be in reasonable agreement. Similar agreements between data and simulation were obtained at the other energies (not shown).

Figure 6 shows the sum of hits in the six layers of the calorimeter for the various energies of the test beam. The data are again compared with the prediction of the simulation. Whereas at the lower energies the agreement between data and simulation is quite good, at the two higher energies the simulation predicts approximately 6% more hits than observed in the data. Based on the following observations, it is believed that the deficiency in the data is due to a loss of efficiency related to the high particle rates in the RPCs: a) the deficiency is observed to be larger for events in the center of the beam spot, b) the deficiency is most noteworthy in the layers with the largest number of *points* (see below), and c) the deficiency is comparatively larger at the end of a spill, as expected due to the exponential decrease in efficiency with time when subjected to high particle fluxes [10].

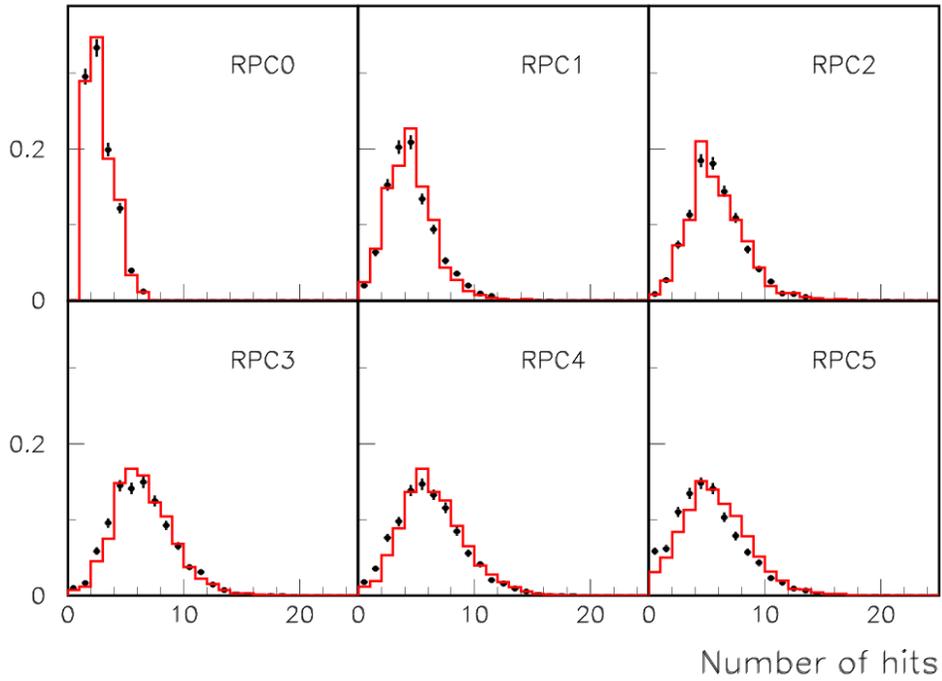

**Figure 4.** Normalized distribution of hits in the six layers of the calorimeter stack from 2 GeV positrons. The red histograms represent the results of a GEANT4 simulation of the set-up together with a standalone simulation of the response of RPCs.

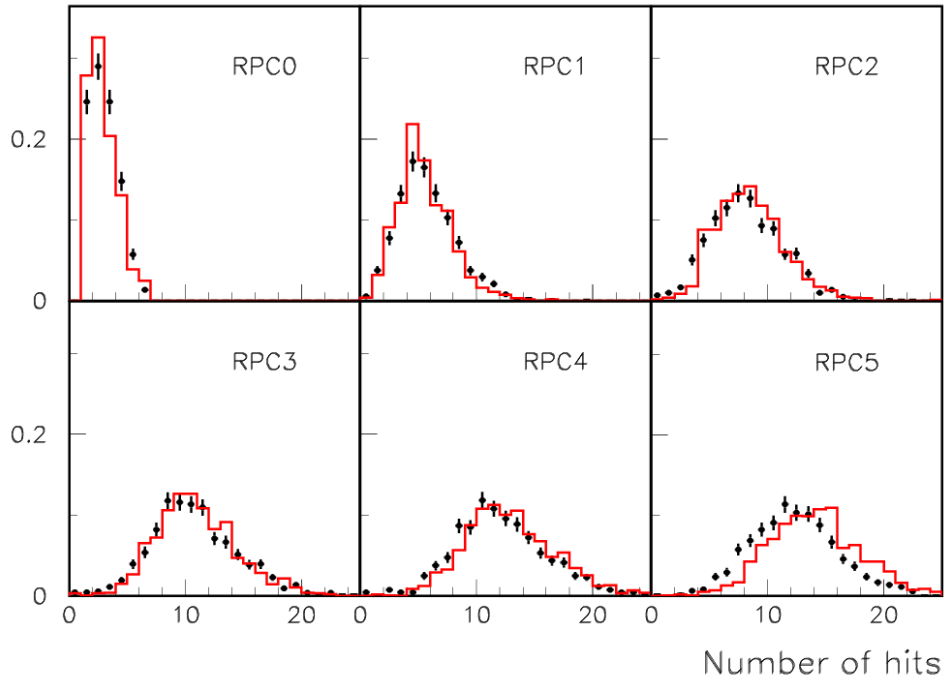

**Figure 5.** Normalized distribution of hits in the six layers of the calorimeter stack from 8 GeV positrons. The red histograms represent the results of a GEANT4 simulation of the set-up together with a standalone simulation of the response of RPCs.

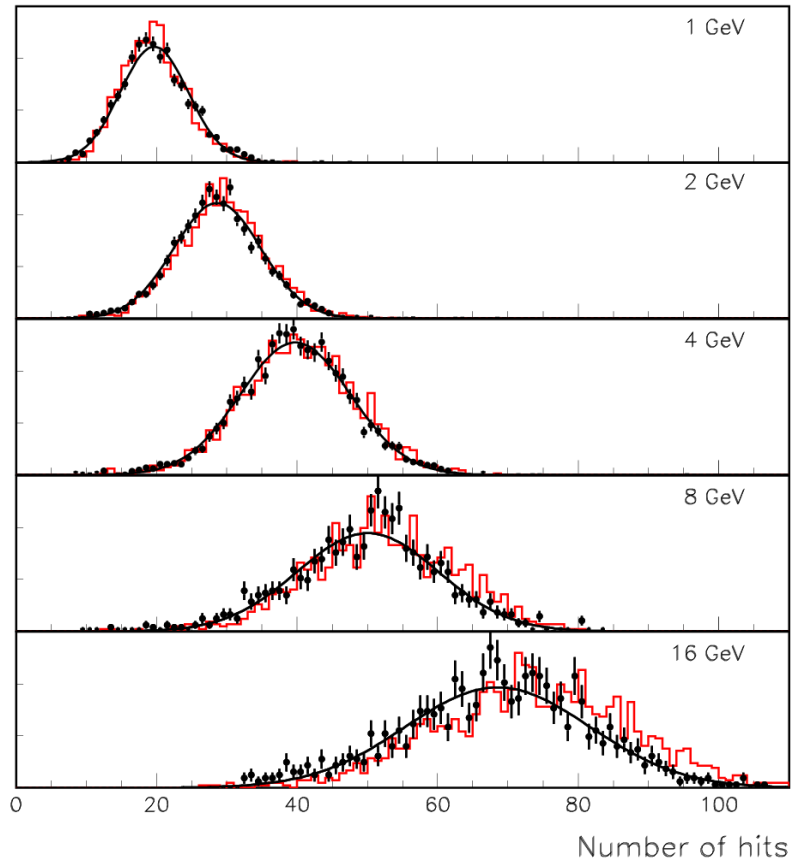

**Figure 6.** Normalized distribution of the sum of hits in the six layers of the calorimeter for 1, 2, 4, 8, and 16 GeV positrons. The red histograms represent the results of a GEANT4 simulation of the set-up together with a standalone simulation of the response of RPCs.

Both the measured and the simulated sum of hits are adequately described by a Gaussian function. The results of the fits to Gaussian distributions are shown in Fig. 7. The means of the distributions are well reproduced, except at the two highest energy points, where the data lie somewhat lower. The response is seen to be highly non-linear due to substantial leakage out the back of the calorimeter and due to the high density of electromagnetic showers and the ensuing overlap of hits on a given pad. Monte Carlo simulations of an extended calorimeter, shown as dashed green curves in the figure, indicate that the first effect is dominant. The measured and simulated widths of the distributions are in reasonable agreement.

Figure 8 shows the average number of hits in individual layers for the various beam energies for both data and simulation. The latter describes the data for all layers at the two lowest energies very well. At the higher energies the simulation predicts larger numbers of hits in the last layers of the stack. The discrepancy is most notable at 16 GeV, where the data show a deficit of about 14% in the last layer. This effect can be explained as due to a loss of efficiency as a result of the high rate of avalanches in the last layers. Also, note that the 16 GeV data starts with a higher number of hits in the first layer compared to the other energies, due to the extra material in the beam line.

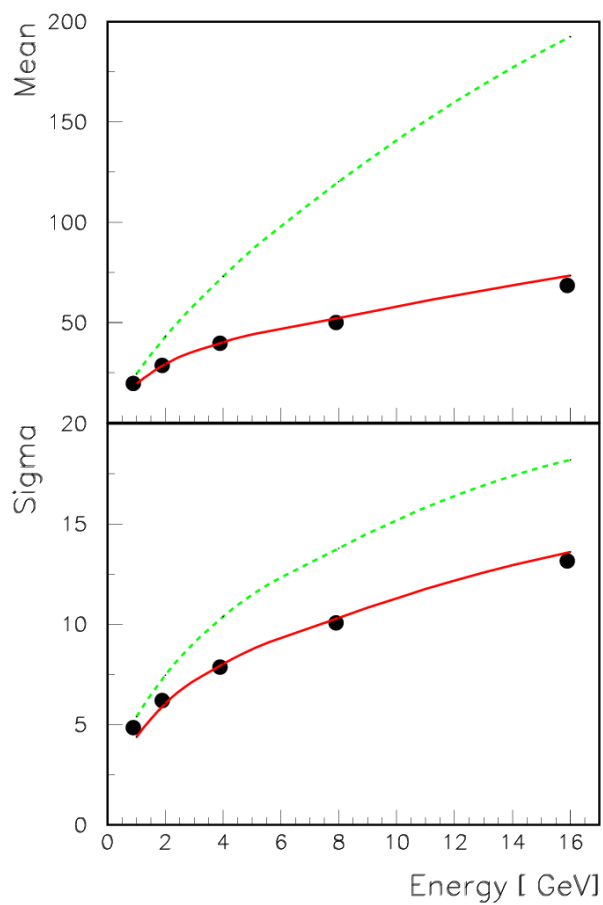

**Figure 7.** Mean and width of the Gaussian fits to the sum of hits in the calorimeter. The red lines represent the results of a GEANT4 simulation of the set-up together with a standalone simulation of the response of RPCs. The dashed green lines indicate the predictions for an infinite calorimeter stack.

In order to measure the lateral shower shapes, the average x and y positions of the hits in a given layer were calculated as their arithmetic mean. These means were fit to a straight line to provide the shower axis for each event. Figure 9 shows for the 2 GeV data the distribution of distances for all hits to their corresponding shower axis. The distributions are adequately reproduced by the simulation, but show a small, systematic depletion at short distances. This effect has been observed at the other beam energies as well and is again due to a loss of efficiency of the chambers at the center of the beam spot. The bin-to-bin variations beyond the statistical significance of the data can be explained as a result of the finite pad size.

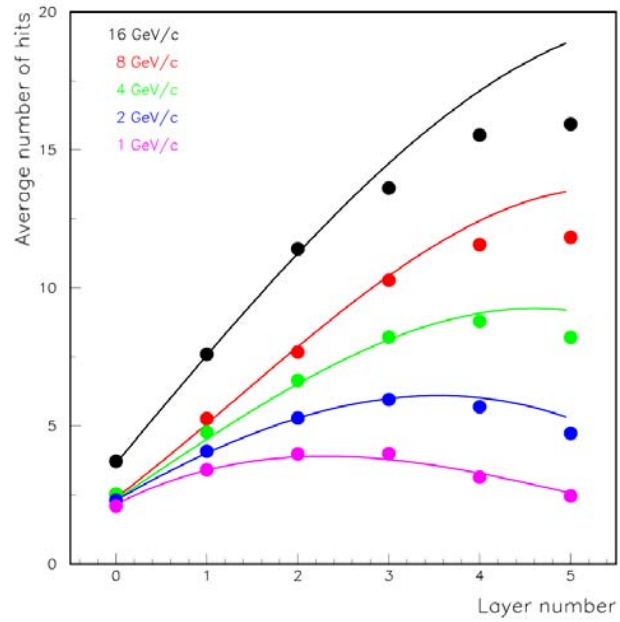

**Figure 8.** Average number of hits as a function of layer number for the various beam energies. The lines represent the results of a GEANT4 simulation of the set-up together with a standalone simulation of the response of RPCs.

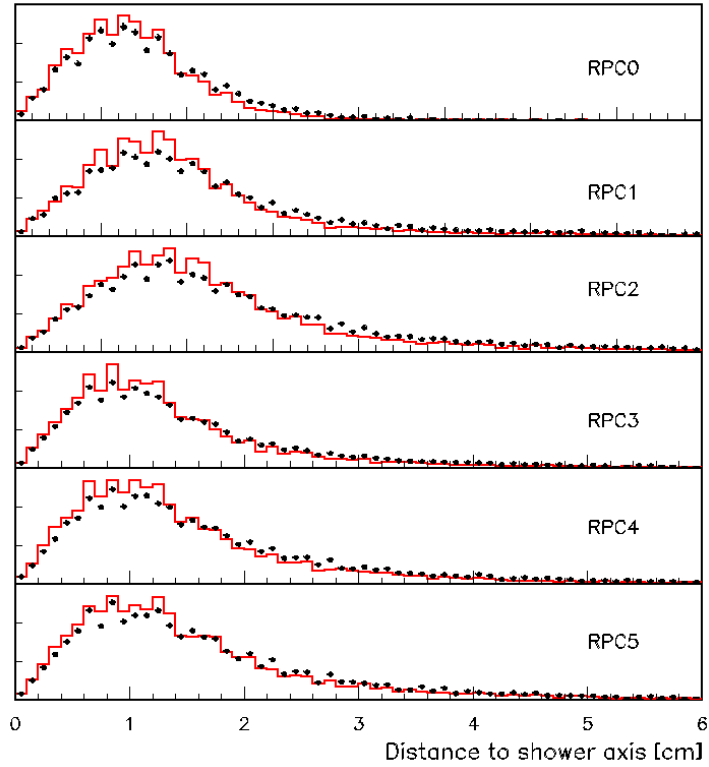

**Figure 9.** Lateral shower shapes for the 2 GeV data. The red line represents the results of a GEANT4 simulation of the set-up together with a standalone simulation of the response of RPCs.

# CONCLUSIONS

A prototype digital hadron calorimeter DHCAL with Resistive Plate Chambers (RPCs) as active elements has been exposed to positrons in the energy range of 1 – 16 GeV. The calorimeter consisted of six layers interleaved with absorber plates with a thickness corresponding to 1.2 $X_0$.

The set-up has been simulated by a GEANT4 based program together with a standalone program to model the response of the RPCs. Three parameters of the response simulation were tuned using data from a broad band muon beam. The last parameter, a short – range distance cut for the efficiency of RPCs, was tuned using the positron data itself.

Detailed measurements of the response of the calorimeter have been presented, including the number of hits/layer, the sum of hits in the entire calorimeter, the average number of hits in each layer, and the lateral shower shape. In general, the simulation reproduces the data quite well. However, some deficiencies of hits in the data are observed, mostly in the high rate regions of the calorimeter. This effect never exceeds 14% and is understood as being due to a loss of efficiency related to high particle fluxes in these regions.

# ACKNOWLEDGEMENTS

We would like to thank the Fermilab test beam crew, in particular Erik Ramberg, Doug Jensen, Rick Coleman and Chuck Brown, for providing us with excellent beam. The University of Texas at Arlington ILC group is acknowledged for providing the two trigger scintillator paddles and their associated trigger logic.